\documentclass[%
preprint,
superscriptaddress,     
 amsmath,amssymb,
 aps,
pra,
]{revtex4-2}

\usepackage{graphicx}
\usepackage{dcolumn}
\usepackage{bm}
\usepackage[usenames,dvipsnames]{color} 
\usepackage[normalem]{ulem} 
\usepackage[dvipsnames]{xcolor}
\usepackage{lineno} 

\def\ldef{\mathrel{\mathop:}=}

\usepackage[linktocpage,colorlinks=true,linkcolor=blue,citecolor=blue,urlcolor=blue,breaklinks=true]{hyperref}   

\begin{document}

\title{Alignment Phase Transition in Socially Driven Motion}

\author{Debasish Sarker}
\affiliation{%
Department of Physics, University of Miami, Coral Gables, Florida 33146, USA}%

\author{Yi Zhang}
\affiliation{%
Department of Physics, University of Miami, Coral Gables, Florida 33146, USA}%

\author{Lynn K. Perry}
\affiliation{%
Department of Psychology, University of Miami, Coral Gables, Florida 33146, USA}%

\author{Daniel S. Messinger}
\affiliation{%
Department of Psychology, University of Miami, Coral Gables, Florida 33146, USA}%

\author{Chaoming Song}
\email{c.song@miami.edu}
\affiliation{%
Department of Physics, University of Miami, Coral Gables, Florida 33146, USA}%

\begin{abstract}
Collective human movement is a hallmark of complex systems, exhibiting emergent order across diverse settings, from pedestrian flows to biological collectives. In high-speed scenarios, alignment interactions ensure efficient flow and navigation. In contrast, alignment in low-speed, socially engaged contexts emerges not from locomotion goals but from interpersonal interaction. Using high-resolution spatial and orientation data from preschool classrooms, we uncover a sharp, distance-dependent transition in pairwise alignment patterns that reflects a spontaneous symmetry breaking between distinct behavioral phases. Below a critical threshold of approximately $0.65$\,m, individuals predominantly align side-by-side; beyond this range, face-to-face orientations prevail. We show that this transition arises from a distance-dependent competition among three alignment mechanisms: parallelization, opposition, and reciprocation, whose interplay generates a bifurcation structure in the effective interaction potential. A Fourier-based decomposition of empirical orientation distributions reveals these mechanisms, enabling the construction of a minimal pseudo-potential model that captures the alignment transition as a non‑equilibrium phase transition. Monte Carlo simulations using the inferred interaction terms closely reproduce the empirical patterns. These findings establish a quantitative framework for social alignment in low-speed human motion, extending active matter theory to a previously unexplored regime of socially mediated orientation dynamics, with implications for modeling coordination and control in biological collectives and artificial swarms.
\end{abstract}
\maketitle
\newpage

\section{Introduction}

Collective human motion exhibits emergent order across spatial and temporal scales, including phase transitions and self-organization, and stands as a paradigmatic system in non‑equilibrium statistical physics~\cite{touchette2009thelarge, ginzburg2009onthe}, particularly within the realm of active matter, characterized by self-driven agents~\cite{marchetti2013hydrodynamics}. These phenomena span from urban transportation and mass migration \cite{rodrigue2024thegeography, castles2013theage}, to pedestrian flows \cite{helbing1995social, montello2005navigation, moussad2010thewalking, feliciani2016empirical}, playground free-play \cite{veiga2018monitoring}, and large-scale crowd dynamics at festivals and public events \cite{silverberg2013collective}, to social alignment patterns in classrooms \cite{zhang2024emergence}. Recent advances in high-resolution tracking technologies have enabled detailed analysis of pedestrian behavior in naturalistic environments, revealing nuanced patterns of movement, orientation, and social interaction \cite{willems2020pedestrian}. These phenomena emerge from local interactions among individuals, yet the rules governing such interactions remain incompletely understood, especially in regimes where movement is not driven by navigation goals but by social interaction.

A large body of work has focused on high-speed collective motion, where individuals prioritize flow efficiency and collision avoidance, such as in marathon races \cite{bain2019dynamic}, bottleneck evacuation \cite{tordeux2020prediction}, pedestrian bidirectional streams \cite{feliciani2016empirical}, or constrained settings like stairways \cite{pouw2024highstatistics}, and align primarily to support efficient navigation \cite{moussad2011howsimple}. In such systems, dominant models attribute alignment to a mechanism of parallelization, where agents locally synchronize their heading direction with neighbors to support efficient locomotion \cite{vicsek1995novel, helbing2005selforganized, couzin2005effective}. This framework has proven powerful in describing flocking, swarming, and crowd flow, and it continues to define the theoretical baseline in the physics of collective behavior \cite{helbing2001traffic}.

However, many natural contexts, including classrooms, conferences, playground free-play \cite{veiga2018monitoring}, and social gatherings, operate in a contrasting regime: low-speed movement punctuated by sustained interpersonal engagement. Here, individuals do not merely coordinate trajectories; they adjust their orientations, modulate personal space, and negotiate social roles \cite{zanlungo2014potential, hall1966thehidden, argyle1965eyecontact}. Despite their ubiquity, such low-speed, interaction-driven systems have received limited attention in physics. Existing alignment models fail to capture the distinct structure of communicative orientation that arises when social connection, rather than locomotion, is the driving force.

Recent empirical evidence suggests that low-speed human motion may constitute a distinct class of active matter. In preschool classrooms, spatial distributions of children reveal coexisting high- and low-density regions, reminiscent of liquid–vapor phase coexistence \cite{zhang2024emergence}. This raises a fundamental question: Do alignment patterns in socially engaged motion also exhibit phase-like transitions, governed not by kinetic alignment, but by social interaction potentials?

Here, we uncover a sharp transition in pairwise alignment patterns that depends systematically on interpersonal distance. Using high-resolution position and orientation data from 89 preschoolers, we find that individuals prefer side-by-side orientation at short range, and shift to face-to-face alignment beyond a critical threshold of approximately $0.65$\,m. Strikingly, this transition cannot be explained by existing models. This transition reflects a spontaneous symmetry breaking between two alignment phases, side-by-side and face-to-face, driven by a bifurcation in the underlying interaction potential. The critical point at which these regimes interchange resembles a non‑equilibrium phase transition in soft active systems. By decomposing the empirical alignment distributions, we identify three competing mechanisms, parallelization, opposition, and reciprocation, whose strengths vary with distance. These components form the basis of a new pseudo-potential model that quantitatively captures the observed symmetry-breaking transition. Our findings reveal alignment as an emergent physical order in low-speed social motion and provide a quantitative bridge between human dynamics, non‑equilibrium statistical physics, and soft active systems.

\section{Results}

\begin{figure*}
  \includegraphics[width=1\linewidth]{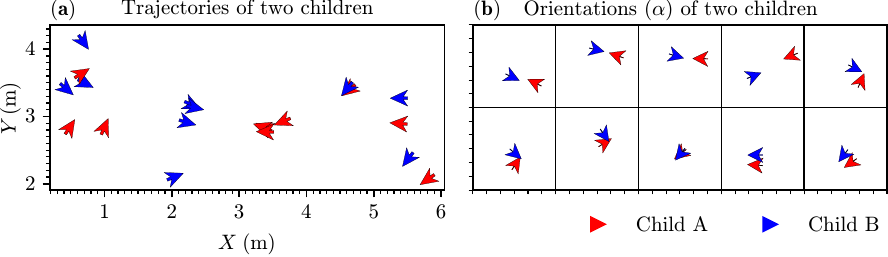}
  \caption{\textbf{Visualizing Orientation Dynamics.} (a) Trajectories of two children across ten randomly selected instances from the LC1 dataset, with arrows indicating their orientation at each time step. (b) Pairwise corresponding time series of interpersonal distance and relative orientation angle (\( \alpha \)) for the same instances. Ticks mark $0.5$\,m intervals.}
  \label{fig:trajectories}
\end{figure*}

A preliminary visualization (Fig.~\ref{fig:trajectories}) illustrates how individuals in a preschool classroom dynamically adjust their body orientations ($\alpha$) in response to changing spatial proximity. When far apart, individuals move largely independently, showing minimal coordination \cite{rio2018local}. As proximity increases, their orientations begin to respond to one another in socially meaningful ways. During joint activities such as group navigation or construction, individuals align side-by-side to facilitate coordination \cite{schmitz2023smaller}. In contrast, moderate distances often correspond to face-to-face alignment, particularly during reciprocal or competitive exchanges \cite{gifford1982projected, hornischer2022modeling}. These observations suggest that interpersonal distance plays a critical role in modulating orientation patterns, consistent with prior work on embodied social interaction \cite{kendon1990conducting}.

To quantify these patterns, we measured relative angles $\theta_1 \ldef \alpha_1 - \alpha_{12}$ and $\theta_2 \ldef \alpha_2 - \alpha_{21}$ for each dyad, where $\alpha_{12}$ and $\alpha_{21}$ represent the bearing angles of relative displacement. These variables characterize each individual's orientation with respect to their partner, enabling dyadic comparison. To assess whether classical alignment mechanisms suffice to explain these observations, we contrast our findings with simulations of the Vicsek model (VM) \cite{vicsek1995novel}, which implements alignment via a parallelization rule.

\begin{figure*}
  \includegraphics[width=1\linewidth]{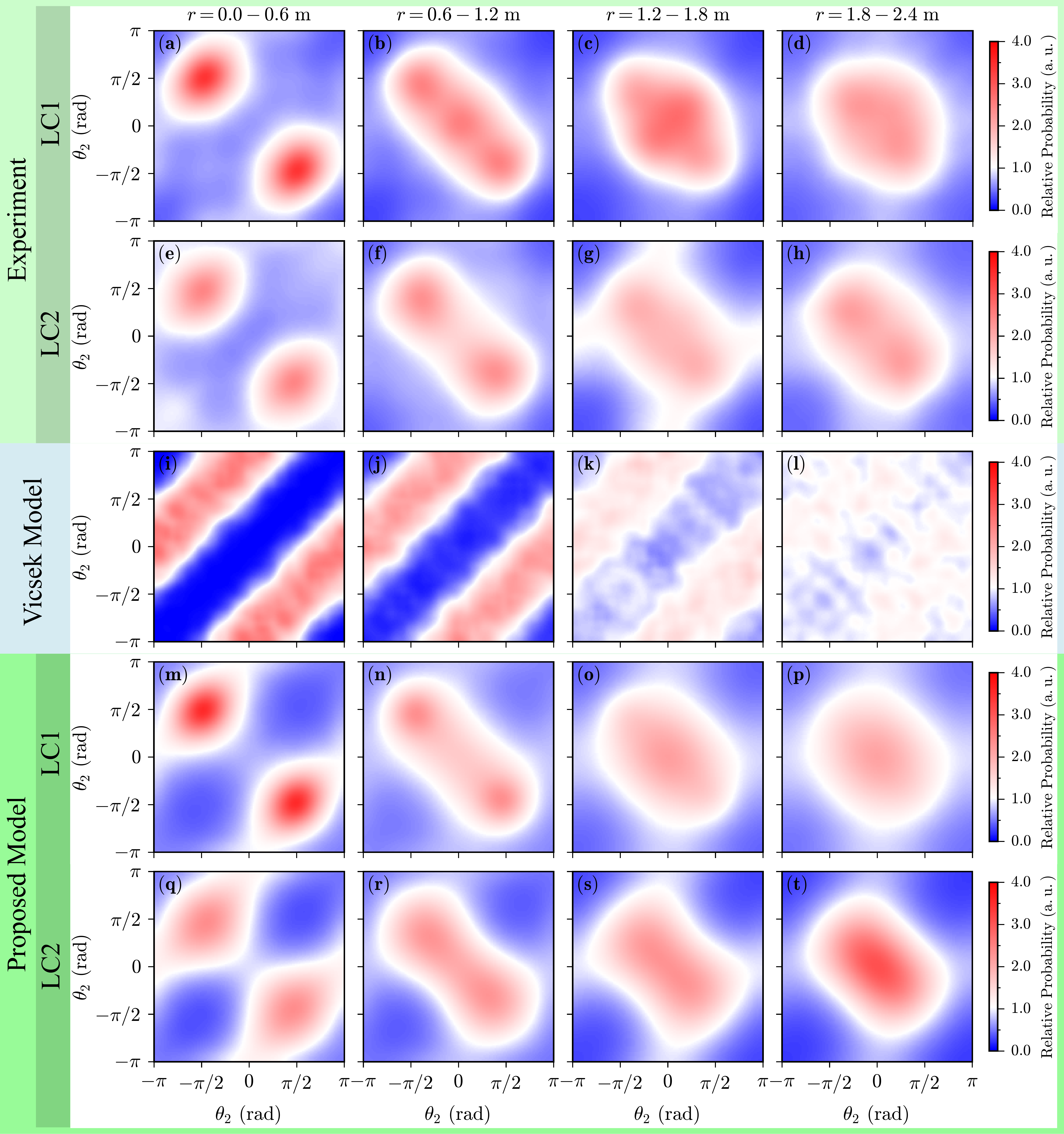}
  \caption{\textbf{Alignment Distributions Across Empirical Data and Models.} Panels (a--d) and (e--h) show empirical pairwise alignment distributions across varying interpersonal distances $r$ from datasets LC1 and LC2, respectively. Panels (i--l) display patterns generated by the standard Vicsek model. Panels (m--p) and (q--t) show heatmaps from Monte Carlo simulations of our proposed model for LC1 and LC2, respectively. White indicates uniform probability; red, elevated; blue, suppressed.}
  \label{fig:alignment_heatmaps}
\end{figure*}

Figure~\ref{fig:alignment_heatmaps} compares the empirical alignment distribution $P(\theta_1, \theta_2)$ (Fig.~\ref{fig:alignment_heatmaps}a--h) with predictions from the Vicsek model (Fig.~\ref{fig:alignment_heatmaps}i--l). In the VM, alignment follows a Gibbs-like distribution,
\begin{equation}
    \ln P(\theta_1,\theta_2) \approx -V(\theta_1,\theta_2),
    \label{eqn:1}
\end{equation}
with pseudo-potential $V_{VM}(\theta_1, \theta_2) = -J\cos(\theta_1 - \theta_2)$ generating a social force $f(\theta_1,\theta_2) = J\sin(\theta_1 - \theta_2)$ \cite{helbing1995social}. Although $V_{VM}$ lacks symmetry due to active matter effects \cite{zanlungo2014potential}, this does not alter the dominant outcome: strong parallel alignment along $\theta_1 - \theta_2 = \pm \pi$. VM predictions remain robust across variations in neighborhood radius $r_0$, speed $v_0$, and angular noise $\eta$, making it an effective benchmark.

In contrast, the empirical data show rich structure beyond VM predictions. As shown in Fig.~\ref{fig:alignment_heatmaps}a--d (LC1) and Fig.~\ref{fig:alignment_heatmaps}e--h (LC2), alignment patterns vary systematically with distance. At short range ($r < 0.6$ m), peaks near $(\pm \frac{\pi}{2}, \mp \frac{\pi}{2})$ indicate strong side-by-side alignment. At moderate distances ($1.2$--$1.8$ m), distributions center near $(0,0)$, reflecting face-to-face orientation (0,0). An intermediate regime ($0.6$--$1.2$ m) displays gradual transition between these two modes. Beyond $3$ m, alignment becomes uniform, indicating absence of coordination \cite{messinger2019continuous}. These findings reveal a previously undocumented, distance-dependent transition in alignment type \cite{Couzin2002CollectiveMemory}.

\begin{figure*}
  \includegraphics[width=1\linewidth]{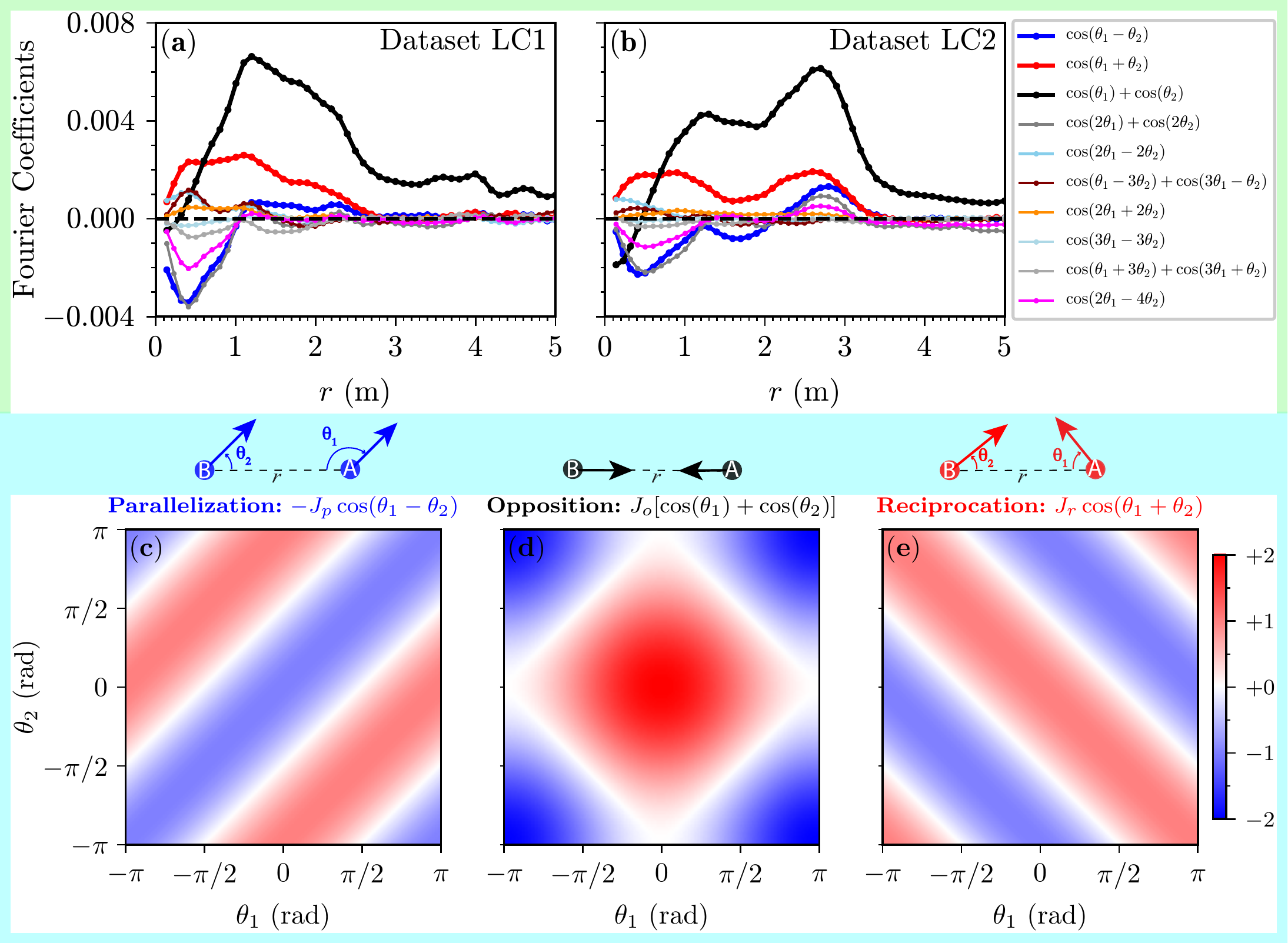}
  \caption{\textbf{Fourier components of alignment as a function of distance.} (a,b) Real-valued cosine Fourier coefficients from datasets LC1 and LC2, respectively, plotted against interpersonal distance $r$. (c--e) Schematic alignment patterns corresponding to the three dominant terms. Red indicates high probability; blue, low.}
  \label{fig:fourier}
\end{figure*}

To explain these transitions, we perform a Fourier expansion:
\begin{equation}
-\ln P(\theta_1, \theta_2) = \sum_{n,m} a_{n,m} e^{i(n\theta_1 + m\theta_2)},
\label{eqn:3}
\end{equation}
extracting spectral components $a_{n,m}$ that dominate alignment behavior. Figure~\ref{fig:fourier}a--b shows the leading real-valued cosine terms from LC1 and LC2. The VM term $\cos(\theta_1 - \theta_2)$ (Fig.~\ref{fig:fourier}c) is substantial at small $r$, confirming parallelization. However, it is not the dominant term overall. The leading contribution arises from $\cos(\theta_1) + \cos(\theta_2)$ (Fig.~\ref{fig:fourier}d), which favors mutual orientation toward the midpoint, defining \emph{opposition}. Also prominent is $\cos(\theta_1 + \theta_2)$ (Fig.~\ref{fig:fourier}e), governing \emph{reciprocation}, where individuals align reflexively. Higher-order terms contribute less and are deferred to the Discussion. Sine terms are negligible.

We integrate these mechanisms into a minimal pseudo-potential model:
\begin{align} \label{eqn:model}
    V(r, \theta_1, \theta_2) &= J_p(r) \cos(\theta_1 - \theta_2)- 2J_o(r) \left(\cos(\theta_1) + \cos(\theta_2)\right) -J_r(r) \cos(\theta_1 + \theta_2) + V_0(r).
\end{align}
where $J_p$, $J_o$, $J_r$ reflect the strengths of parallelization, opposition, and reciprocation as functions of $r$, and $V_0(r)$ normalizes marginals. Each term corresponds to a specific alignment mechanism.

Monte Carlo simulations \cite{newman1999monte} using Eq.~\eqref{eqn:model} yield heatmaps that closely reproduce the empirical alignment patterns across LC1 and LC2 (Fig.~\ref{fig:alignment_heatmaps}m--t), demonstrating the predictive sufficiency of the model.

To characterize this transition, we analyze the stationary points of $V(r, \theta_1, \theta_2)$ (see Methods). The critical configuration depends on the sign of $\Delta J = J_o - J_p$. When $\Delta J < 0$, parallelization dominates, yielding a side-by-side phase (Fig.~\ref{theory_simulation}a). When $\Delta J > 0$, opposition prevails, producing face-to-face alignment (Fig.~\ref{theory_simulation}c). At $\Delta J = 0$, both configurations are degenerate (Fig.~\ref{theory_simulation}b).

\begin{figure}
  \includegraphics[width=1\linewidth]{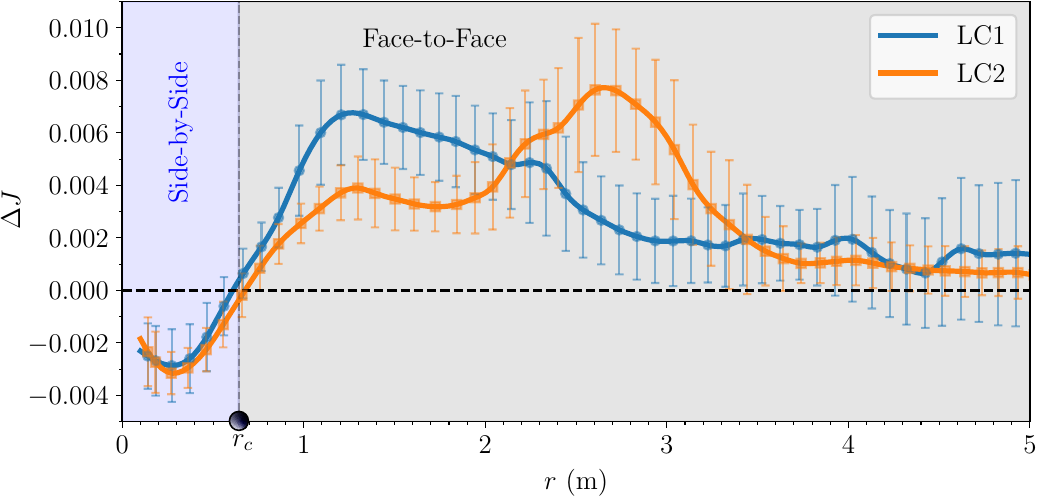}
  \caption{\textbf{Phase diagram.} Distance-dependent difference in opposition and parallelization strength $\Delta J = J_o - J_p$. A transition from side-by-side (blue) to face-to-face (gray) occurs at $r_c \approx 0.65$\,m, where $\Delta J = 0$.}
  \label{fig:phase-diagram}
\end{figure}

Figure~\ref{fig:phase-diagram} shows $\Delta J(r)$ from empirical data. A crossover at $r_c \approx 0.65$\,m confirms the predicted transition, robust across classrooms with different layouts and movement patterns. This agreement between data and theory establishes a distance-dependent symmetry-breaking transition between alignment phases, governed by competing social interaction mechanisms.

\section{Discussion}

Our findings reveal a distance-dependent transition in alignment patterns governed by the interplay of competing social mechanisms. At short distances, \emph{parallelization} dominates, favoring side-by-side alignments that support joint action and spatial coordination. Beyond a critical threshold ($r > 0.65$\,m), \emph{opposition} becomes dominant, driving face-to-face alignments that facilitate direct social engagement \cite{ransom2022facetoface, wang2022close}. This transition reflects a fundamental trade-off between coordination and communication \cite{moussad2010thewalking, gregorj2023social}, with the crossover marking a balance point between the two. Strikingly, this behavior recurs consistently across datasets and classroom contexts, suggesting a robust organizing principle underlying low-speed social alignment, akin to threshold transitions observed in many-body systems \cite{touchette2009thelarge}.

The minimal interaction model introduced here not only reproduces the empirically observed transition but also establishes a foundation for future theoretical generalizations. For instance, the current framework does not explicitly incorporate higher-order social dynamics, such as status asymmetries or group-level structure, which may shape alignment preferences in larger or more hierarchical collectives. Extending the model to capture such effects could broaden its applicability to complex social environments. Additionally, the identified interaction terms, parallelization, opposition, and reciprocation, could support richer phase diagrams when combined with dynamical features like clustering, task switching, or topological constraints, as hinted at in recent empirical work \cite{zhang2024emergence}. While our Fourier-based model emphasizes dominant harmonics, incorporating higher-order alignment terms could reveal subtle orientation behaviors shaped by sociocultural norms or environmental layout. Future work could also investigate how factors such as group density, classroom design, or activity context modulate alignment regimes, offering a path toward unified models of socially guided motion. The underlying potential landscape exhibits bifurcation behavior with symmetry-breaking minima, placing this system in correspondence with canonical models of second-order phase transitions~\cite{cross1993pattern, chaikin1995principles, kardar2007collective}.

Taken together, our results advance the understanding of low-speed, socially engaged human movement as a non‑equilibrium physical system with emergent alignment phases. The framework we develop bridges microscopic orientation dynamics with macroscopic collective behavior, motivating a shift from navigation-centric to interaction-centric models of human motion. Our framework shares structural parallels with Ginzburg–Landau models of continuous symmetry breaking~\cite{ginzburg2009onthe}, but differs in its non-Hamiltonian interaction terms and explicit distance dependence, hallmarks of non‑equilibrium active systems. The resulting binary phase structure, side-by-side versus face-to-face alignment, also echoes the behavior of spin models with competing interactions~\cite{ising1925beitrag, stanley1971phase}, grounding our findings in familiar symmetry classes while extending them to socially driven motion. This perspective holds broad interdisciplinary relevance, from developmental science and pedagogy to crowd safety, epidemic modeling \cite{zhang2022simulating}, and swarm robotics \cite{repiso2020adaptive}. By quantifying the microscopic rules of social alignment, this work opens new directions for modeling interactive agents in physical and artificial systems alike.

\section{Methods}

\subsection{Data collection}
We collected high-resolution spatial and orientation data from two preschool classrooms (LC1 and LC2) using ultra-wideband radio-frequency identification (UWB-RFID) sensors, operating at a sampling rate of 2--4\,Hz. Each child wore a custom vest equipped with two spatially separated tags, one on each shoulder, enabling precise, real-time reconstruction of both position and facing direction. This dual-tag configuration allowed us to resolve heading vectors without reliance on external pose estimation. Over a four-year period, we obtained 236 hours of movement and orientation recordings from 89 children, providing a rich dataset for characterizing low-speed social alignment dynamics (see SI Table~S1). Technical specifications, calibration protocols, environmental constraints, and preprocessing methods are detailed in the Supplementary Material ~\cite{messinger2022chapter, cattuto2010dynamics}.

\subsection{Theoretical analysis}

To analytically investigate the alignment transition revealed in empirical data, we examine the pseudo-potential function \( V(r, \theta_1, \theta_2) \) introduced in Eq.~\eqref{eqn:model}, where \( \theta_1 \) and \( \theta_2 \) denote the relative orientations of two interacting individuals. The equilibrium configurations of the system correspond to stationary points of this potential, satisfying the conditions:
\begin{subequations}
\begin{align}
    \partial_{\theta_1} V &= J_{r}\sin(\theta_1 + \theta_2) - J_{p}\sin(\theta_1 - \theta_2) + 2J_{o} \sin(\theta_1) = 0, \label{eq3} \\
    \partial_{\theta_2} V &= J_{r}\sin(\theta_1 + \theta_2) + J_{p}\sin(\theta_1 - \theta_2) + 2J_{o} \sin(\theta_2) = 0. \label{eq4}
\end{align}
\end{subequations}
Here, \( J_p \), \( J_o \), and \( J_r \) are the distance-dependent coupling strengths corresponding to parallelization, opposition, and reciprocation, respectively.

These coupled equations yield two analytically tractable solutions:
\begin{itemize}
    \item A \textit{trivial solution} at \( \theta_1 = \theta_2 = 0 \), corresponding to face-to-face alignment.
    \item A \textit{non-trivial solution} at \( \theta_1 = -\theta_2 \), with \( \cos(\theta_1) = \cos(\theta_2) = J_o / J_p \), representing side-by-side alignment. This solution is physically admissible only when \( J_o \leq J_p \), consistent with the idea that opposition cannot dominate over parallelization in the side-by-side regime.
\end{itemize}

\begin{figure*}
  \includegraphics[width=1\linewidth]{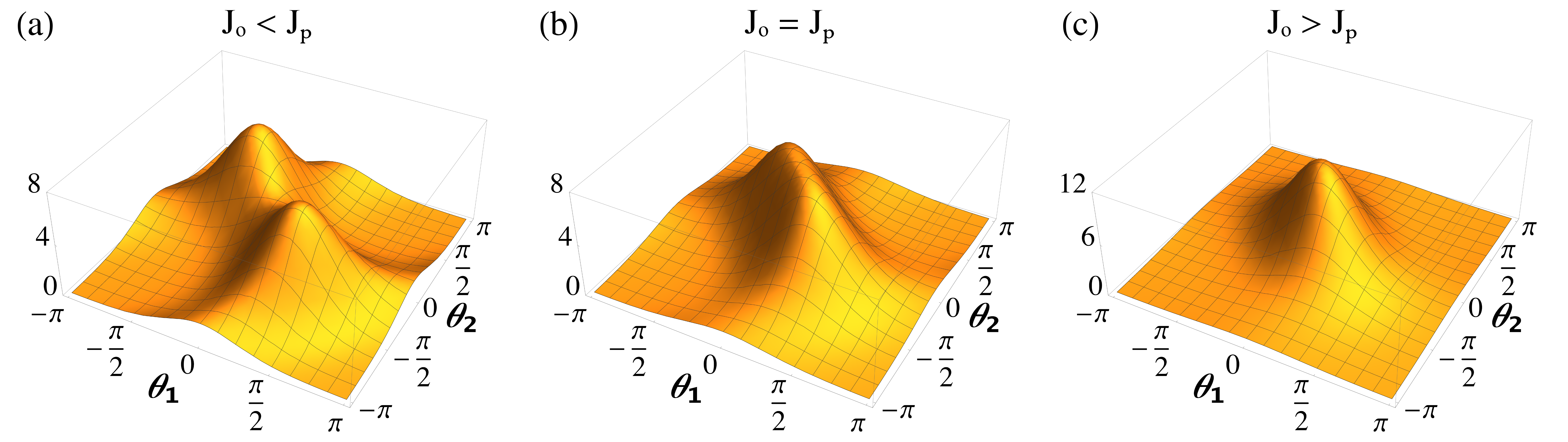}
  \caption{\textbf{Theoretical Distributions.} (a) Side-by-side regime: \( J_o < J_p \) yields two maxima at opposing quadrants. (b) Critical transition: \( J_o = J_p \) results in a degenerate coexistence. (c) Face-to-face regime: \( J_o > J_p \) leads to a single global minimum at the center. These analytic configurations match the empirical alignment transitions (Fig. \ref{fig:alignment_heatmaps}a-h).}
  \label{theory_simulation}
\end{figure*}

To determine which of these configurations represents a stable minimum, we compute the Hessian matrix of the pseudo-potential evaluated at \( \theta_1 = \theta_2 = 0 \):
\begin{equation}
H(0,0) = \begin{pmatrix}
J_{r} - J_{p} + 2J_{o} & J_{r} + J_{p} \\
J_{r} + J_{p} & J_{r} - J_{p} + 2J_{o}
\end{pmatrix}.
\end{equation}
The eigenvalues of this matrix are $\lambda_1 = 2(J_{r} + J_{o})$ and $\lambda_2 = 2(J_{o} - J_{p})$ which define the local curvature of the potential and hence the stability of the solution. This yields three qualitative regimes:
\begin{enumerate}
    \item[\textbf{i)}] \textbf{Side-by-side regime} (\( J_o < J_p \)): Here, \( \lambda_2 < 0 \), and the trivial solution becomes a saddle point. The stable configuration lies at \( \theta_1 = -\theta_2 \), resulting in two off-center maxima (Fig.~\ref{theory_simulation}a).
    \item[\textbf{ii)}] \textbf{Transition point} (\( J_o = J_p \)): At this critical point, \( \lambda_2 = 0 \), and both alignment modes are equally likely, giving rise to a degenerate equilibrium (Fig.~\ref{theory_simulation}b).
    \item[\textbf{iii)}] \textbf{Face-to-face regime} (\( J_o > J_p \)): When opposition dominates, both eigenvalues are positive, and the system minimizes its potential at \( \theta_1 = \theta_2 = 0 \), leading to a central peak (Fig.~\ref{theory_simulation}c).
\end{enumerate}

This analytical phase structure aligns closely with our empirical findings and Monte Carlo simulations (Figs.~\ref{fig:alignment_heatmaps}a-h and \ref{fig:phase-diagram}), confirming that the transition from side-by-side to face-to-face alignment results from a symmetry-breaking bifurcation in the effective interaction landscape. Full derivations and extensions, including higher-order terms, are provided in the Supplementary Material. While this study focuses on preschool classrooms, the proposed framework applies more broadly to any system of interacting agents exhibiting distance-dependent alignment under low-speed constraints, including robotic swarms, pedestrian clusters, or animal collectives.

\section*{Author Contributions}
D. Sarker and Y. Zhang contributed equally to the work.



\begin{acknowledgments}
This work was partially supported by the National Science Foundation under Grants No. 2150830 and IBSS-1620294; the Institute of Education Sciences under Grant No. R324A180203; the National Institutes of Health under Grant No. R01DC018542; and the Simons Foundation Autism Research Initiative under Grant No. SFI-AR-HUMAN-00004115-01.
\end{acknowledgments}

\bibliography{ref_verified}

\end{document}